\newif\iffigs\figstrue

\documentstyle[12pt]{article}
\iffigs
  \input epsf
\else
\fi

\batchmode
  \newfont{\footscrfont}{rsfs10}
  \newfont{\footbbbfont}{msbm10}
\errorstopmode

\newif\ifscrf\scrftrue
\ifx\footscrfont\nullfont
  \scrffalse
\fi

\newif\ifamsf\amsftrue
\ifx\footbbbfont\nullfont
  \amsffalse
\fi

\def\ppnumber{\vbox{\baselineskip14pt\hbox{CLNS-96/1409}
\hbox{hep-th/9605131}}}
\def\ppdate{May 1996}
\def\pplogo{\vbox{\kern-\headheight\kern -15pt
\halign{##&##\hfil\cr&{
\ppnumber}\cr\rule{0pt}{2.5ex}&\ppdate\cr}
}}

\makeatletter
\date{}
\def\dedicatory#1{\def\@date{\normalsize\it#1}}
\def\subjclass#1{\def\@thefnmark{}\@footnotetext{1991
    {\it Mathematics Subject Classification.} #1}}
\def\keywords#1{\def\@thefnmark{}\@footnotetext{
    {\it Key words and phrases.} #1}}

\def\ps@firstpage{\ps@empty \def\@oddhead{\hss\pplogo}%
  \let\@evenhead\@oddhead 
}
\def\maketitle{\par
 \begingroup
 \def\thefootnote{\fnsymbol{footnote}}
 \def\@makefnmark{\hbox
 to 0pt{$^{\@thefnmark}$\hss}}
 \if@twocolumn
 \twocolumn[\@maketitle]
 \else \newpage
 \global\@topnum\z@ \@maketitle \fi\thispagestyle{firstpage}\@thanks
 \endgroup
 \setcounter{footnote}{0}
 \let\maketitle\relax
 \let\@maketitle\relax
 \gdef\@thanks{}\gdef\@author{}\gdef\@title{}\let\thanks\relax}

\def\abstract{\if@twocolumn
\section*{Abstract}
\else \small
\begin{center}
{\bf ABSTRACT}
\end{center}
\quotation
\fi}

\def\thebibliography#1{\section*{References\@mkboth
 {REFERENCES}{REFERENCES}}\small\list
 {[\arabic{enumi}]}{\settowidth\labelwidth{[#1]}\leftmargin\labelwidth
 \advance\leftmargin\labelsep
 \usecounter{enumi}}
 \def\newblock{\hskip .11em plus .33em minus .07em}
 \sloppy\clubpenalty4000\widowpenalty4000
 \sfcode`\.=1000\relax}

\newif\iffn\fnfalse

\@ifundefined{reset@font}{\let\reset@font\empty}{} 
\long\def\@footnotetext#1{\insert\footins{\reset@font\footnotesize
    \interlinepenalty\interfootnotelinepenalty
    \splittopskip\footnotesep
    \splitmaxdepth \dp\strutbox \floatingpenalty \@MM
    \hsize\columnwidth \@parboxrestore
   \edef\@currentlabel{\csname p@footnote\endcsname\@thefnmark}\@makefntext
    {\rule{\z@}{\footnotesep}\ignorespaces
      \fntrue#1\fnfalse\strut}}}

\makeatother

\ifamsf
  \newfont{\bigbbbfont}{msbm10 scaled\magstep2}
  \newfont{\bbbfont}{msbm10 scaled\magstep1}  
  \newfont{\smallbbbfont}{msbm8}
  \newfont{\tinybbbfont}{msbm6}
  \newfont{\smallfootbbbfont}{msbm7}
  \newfont{\tinyfootbbbfont}{msbm5}
\fi

\ifscrf
  \newfont{\scrfont}{rsfs10 scaled\magstep1}  
  \newfont{\smallscrfont}{rsfs7}
  \newfont{\tinyscrfont}{rsfs7}
  \newfont{\smallfootscrfont}{rsfs7}
  \newfont{\tinyfootscrfont}{rsfs7}
\fi

\ifamsf
  \newcommand{\Bbb}[1]{\iffn
      \mathchoice{\mbox{\footbbbfont #1}}{\mbox{\footbbbfont #1}}
      {\mbox{\smallfootbbbfont #1}}{\mbox{\tinyfootbbbfont #1}}\else
      \mathchoice{\mbox{\bbbfont #1}}{\mbox{\bbbfont #1}}
      {\mbox{\smallbbbfont #1}}{\mbox{\tinybbbfont #1}}\fi}
\else
  \def\bigbbbfont{\bf}
  \def\Bbb{\bf}
\fi

\ifscrf
  \newcommand{\Scr}[1]{\iffn
    \mathchoice{\mbox{\footscrfont #1}}{\mbox{\footscrfont #1}}
    {\mbox{\smallfootscrfont #1}}{\mbox{\tinyfootscrfont #1}}\else
    \mathchoice{\mbox{\scrfont #1}}{\mbox{\scrfont #1}}
    {\mbox{\smallscrfont #1}}{\mbox{\tinyscrfont #1}}\fi}
\else
  \def\Scr{\cal}
\fi

\def\operatorname#1{\mathop{\rm #1}\nolimits}

\def\P{{\Bbb P}}

\def\Z{{\Bbb Z}}

\def\Pic{\operatorname{Pic}}
\def\disc{\operatorname{disc}}

\def\SO{\operatorname{SO}}
\def\SU{\operatorname{SU}}
\def\Sp{\operatorname{Sp}}

\def\opeq#1{\advance\lineskip#1 \advance\baselineskip#1
        \advance\lineskiplimit#1}

\def\eqalign#1{\null\,\vcenter{\opeq{2.5\jot}\mathsurround=0pt
        \everycr={}\tabskip=0pt
        \halign{\strut\hfil$\displaystyle{##}$&$\displaystyle{{}##}$\hfil
        \crcr#1\crcr}}\,\null}

\def\CY{Calabi--Yau}

\def\cM{{\Scr M}}

\def\cD{{\Scr D}}

\def\HF#1{{\bf F}_#1}

\def\cMc{{\hfuzz=100cm\hbox to 0pt{$\;\overline{\phantom{X}}$}\cM}}
\def\barcD{{\hfuzz=100cm\hbox to 0pt{$\;\overline{\phantom{X}}$}\cD}}

\def\ff#1#2{{\textstyle\frac{#1}{#2}}}

\ifamsf

\else

\fi

\begin{document}
\setcounter{page}0
\title{\LARGE The SO(32) Heterotic String on a K3 Surface\\[10mm]}
\author{
Paul S. Aspinwall\\[0.7cm]
\normalsize F.R.~Newman Lab.~of Nuclear Studies,\\
\normalsize Cornell University,\\
\normalsize Ithaca, NY 14853\\[10mm]
Mark Gross\\[0.7cm]
\normalsize Department of Mathematics,\\
\normalsize Cornell University,\\
\normalsize Ithaca, NY 14853\\[5mm]
}

{\hfuzz=10cm\maketitle}

\def\Large{\large}
\def\LARGE{\large\bf}

\vskip 1cm

\begin{abstract}

The $\SO(32)$ heterotic string on a K3 surface is analyzed in terms of
the dual theory of a type II string (or F-theory) on an elliptically
fibred \CY\ manifold. The results are in beautiful agreement with
earlier work by Witten using very different methods. In particular, we
find gauge groups of $\SO(32)\times\Sp(k)$ appearing at points in the
moduli space identified with point-like instantons and see
hypermultiplets in the $(\mbox{\bf 32},\mbox{\boldmath {\bf 2}$k$})$
representation becoming massless at the same time.
We also discuss some aspects of the $E_8\times E_8$ case.

\end{abstract}

\vfil\break


\section{Introduction}

There are two heterotic superstring theories in ten dimensions which
give rise to gauge groups $E_8\times E_8$ and $\SO(32)$ respectively. The
simplest description of duality for these two theories looks quite
different. The $E_8\times E_8$ string is believed to be dual to
M-theory compactified on a finite line segment \cite{HW:E8M} whereas
the $\SO(32)$ case is supposedly dual to a type I superstring 
\cite{W:dyn,PW:TypeI}. 

When the heterotic string is compactified on a circle (and therefore
any manifold with a circle as a factor), the two theories become
equivalent in the sense that the $E_8\times E_8$ string compactified
on such a manifold is identical to the $\SO(32)$ case compactified on the
same manifold with different moduli \cite{N:torus,Gins:torus}. It is
interesting to ask whether this holds for other compactification
spaces.  

The case of the $\SO(32)$ string compactified on a K3 surface was
studied in \cite{W:small-i}. There it was shown that, although the
unbroken gauge group was generically $\SO(8)$, by shrinking down some of
the instantons one could achieve non-perturbative $\Sp(1)$ extra gauge
group factors.\footnote{There is a closely-related ambiguity of a
factor of 2 in counting supersymmetries in 5, 6 or 7 dimensions and whether
symplectic groups are denoted $\Sp(k)$ or $USp(2k)$. We will count
quaternionically in both cases and so will be concerned with $N=1$
supersymmetry 
in 6 dimensions with gauge groups containing $\Sp(k)$.}
 If all 24 instantons are shrunk down then the gauge
group is generically $\SO(32)\times\Sp(1)^{24}$. One may further
enhance the gauge group by allowing these point-like instantons to
coalesce. When $k$ instantons coalesce at a generic point in the K3
surface, the $\Sp(1)^k$ factor is expected to be replaced by
$\Sp(k)$. Thus one may obtain $\SO(32)\times\Sp(24)$ as a gauge group
in the case that all 24 instantons coalesce.
The analysis of \cite{W:small-i} and subsequent analysis (see, for
example \cite{GP:open}) relied on this heterotic string's duality with
the type I string, and $D$-brane technology.

The $E_8\times E_8$ string compactified on a K3 surface has recently
been analyzed in \cite{MV:F,MV:F2,SW:6d}. In this case the analysis
has been done in terms of M-theory or F-theory which, for purposes of
this letter, should be thought of as limits of the type II
superstring. In this case, when the instantons are shrunk down to zero
size, no further gauge group appears but instead extra massless ``tensor''
multiplets appear which have no perturbative description in terms of
the heterotic string. 

It was conjectured in \cite{MV:F} that the $\SO(32)$ heterotic string
compactified on a K3 surface is the same as a particular case of the
$E_8\times E_8$ heterotic string compactified on another K3 surface
and so the situation for the two heterotic strings compactified on a K3
surface should be analogous to that of the circle above.

If this is the case we should be able to reproduce the analysis of the
$\SO(32)$ string as in \cite{W:small-i} in the same language as the
$E_8\times E_8$ string was studied in \cite{MV:F,MV:F2}. This is the
goal of this letter. As well as providing yet another highly
non-trivial test of string duality, we believe that the elliptic
fibration language used to describe the heterotic string below will
prove to be the more powerful (although this may be a question of taste).

While completing this work we became aware of \cite{BKV:enhg} which
has some overlap with the results presented here.


\section{The Elliptic Fibration}  \label{s:ell}

In this section we will review the analysis we require from
\cite{MV:F,MV:F2}. This was originally cast in the form of F-theory
but here we choose to rephrase things to enable us to make some
slightly stronger statements.

Begin with the widely-believed proposition that the type IIA string
compactified on a K3 surface is equivalent to the ($E_8\times E_8$ or
$\SO(32)$) heterotic string compactified on a four-torus. Now use this
duality fibre-wise as proposed in \cite{VW:pairs} to generate dual
pairs in the form of a type IIA string compactified on a \CY\
threefold and a heterotic string compactified on K3$\times T^2$ of the
type proposed in \cite{KV:N=2,FHSV:N=2}. The \CY\ threefold is of the
form of a K3-fibration for such a scheme to work but this must be the
case if the weakly-coupled heterotic string can be understood in terms
of a perturbative sigma-model on the \CY\ \cite{AL:ubiq}.

Since we need to handle various K3 surfaces and various fibrations, let
us fix some notation. Let the type IIA string be compactified on the \CY\
threefold $X$. We then have a fibration
\begin{equation}
  X\to\P^1(w),
\end{equation}
with generic K3 fibre, $S_w$, where $w$ is an affine coordinate on the
base $\P^1$. String-string duality then replaces $S_w$ by $T^4_w$ for
the heterotic string. Now let this $T^4_w$ factorize into $T^2_0\times
T^2_w$, where $T^2_0$ is constant over $\P^1(w)$. $T^2_w$ is then the
fibre for another fibration
\begin{equation}
  S_H\to\P^1(w),
\end{equation}
where $S_H$ is a K3 surface such that the heterotic string is
compactified on $S_H\times T^2_0$

Now let the $T^2_0$ of the $S_H\times T^2_0$ acquire infinite area. This will
decompactify the theory to the desired heterotic string compactified
on the K3 surface, $S_H$. If we can understand in terms of the type II
string what this large area limit of $T^2$ corresponds to, then we
will have some type II picture of the heterotic string compactified on
a K3 surface.

We thus regard the N=1 theory in six dimensions of a heterotic string
compactified on a K3 surface as dual to a special limit of a type IIA string
theory compactified on some \CY\ space, $X$. We need to isolate the
$T^2_0$ degrees of freedom from the rest of the theory to allow it to
go to infinite 
size. We thus demand that within $T^4_w$, we have a heterotic string
compactified on $T^2_0\times T^2_w$, where the metric factorizes and no
Wilson lines are wrapped around $T^2_0$. The Narain moduli
space of a heterotic string on $T^4_w$ is thus restricted as
\begin{equation}
  \frac{O(4,20)}{O(4)\times O(20)} \supset \frac{O(2,18)}{O(2)\times
	O(18)}\times\frac{O(2,2)}{O(2)\times O(2)}.
\end{equation}
For the discrete identifications within this space we split the even
self-dual lattice $\Gamma_{4,20}$ into two self-dual lattices 
$\Gamma_{2,18}\oplus\Gamma_{2,2}$.

Such a factorization of the moduli space of heterotic strings on a
torus is dual to a type IIA string compactified on a K3 surface which
has a chosen algebraic type (see, for example, \cite{me:en3g}). 
The factors of the moduli space then correspond to deformation of
complex structure and the deformation of complexified K\"ahler form of
the K3 surface.
In the
case at hand we want the K3 surface to have Picard lattice
$\Gamma_{1,1}$. It follows that this K3
surface must be of the form of an elliptic fibration with a unique global
section.

What we have shown therefore is that if we can decompactify the
heterotic string to 6 dimensions as above, then $X$ must be a K3
fibration where the generic K3 fibre, $S_w$, is itself an elliptic
fibration. We denote this fibration
\begin{equation}
  S_w\to\P^1_w(z),
\end{equation}
where the generic fibre is an elliptic curve $e(w,z)$. The notation
$\P^1_w(z)$ means that the affine coordinate on this $\P^1$ is $z$, and
there are a set of such $\P^1$'s parametrized by $w$.
The fact that this elliptic fibration has a unique section can then be used
to rewrite $X$ itself as an elliptic fibration over some complex
surface. This surface will be of the form of a fibration over $\P^1(w)$
with generic fibre $\P^1_w(z)$. That is, it is the Hirzebruch surface $\HF
n$. It also follows that this elliptic fibration will have a section.

What exactly happens to $X$ as the $T^2$ in the heterotic string goes
off to infinite size? Since the moduli space of the heterotic string
on $T^4$ can be matched point-for-point with the moduli space of a
type IIA string on a K3 surface \cite{AM:K3p}, one can show that the
effect on the K3 fibre within $X$ is as follows. The base $\P^1_w(z)$
blows up to infinite size but the area of the fibre $e(w,z)$ may be
held constant. Also, if we want to make the dilaton of the six
dimensional theory finite, the four-dimensional heterotic string must
go to a weakly-coupled limit as $T^2_0$ gets larger. In terms of $X$,
this means that the base $\P^1(w)$ of $X$ as a K3-fibration goes to
infinite size \cite{AL:ubiq}.

At first sight it looks therefore as if $X$ grows in a way compatible
with the elliptic fibration picture where the fibre's area remains constant
but the base $\HF n$ goes to infinite volume. However, we need to make sure
we are capturing the right degrees of freedom to describe
six-dimensional physics. To do this, one may use M-theory type
arguments along the lines of \cite{W:dyn} or the F-theory arguments of
\cite{Vafa:F} to see that what we should really do is to rescale the
metric on $X$ so that it is the base $\HF n$ that remains finite sized
but the elliptic fibre shrinks down to nothing.

The claim from \cite{MV:F} which we require that we have outlined above
is thus the following. Any heterotic string compactified on a K3
surface is dual to a type IIA string theory compactified on a \CY\
$X$, where $X$ is an elliptic fibration over $\HF n$ with a
section. We also only consider degrees of freedom present when the
fibre is shrunk down to zero size.

Hirzebruch surfaces $\HF n$ are classified by the number $n$. This is
defined such that the self-intersection of the ``exceptional'' rational
curve $C_0\subset\HF n$ is $-n$. We need to know how $n$ is determined
by the heterotic string. This then determines what $X$ is. Recall that
to obtain an anomaly-free theory, the heterotic string is compactified
on a K3 surface along with a principal $G$-bundle, $E$, where $G$ is
either $E_8\times E_8$ or $\SO(32)$ according to which heterotic
string is used and $c_2(E)=24$. The homotopy class of a $G$-bundle over
a K3 surface is defined by a map from $\pi_3(G)$ into $H^4({\rm K3})$
(see, for example, \cite{W:tools}). In the case of $G\cong\SO(32)$ we
have that $\pi_3\cong\Z$ and this map is essentially fixed by the
$c_2$ condition. In the case of $G\cong E_8\times E_8$ however, we
have $\pi_3\cong\Z\oplus\Z$ and there remains an integer degree of
freedom. The degree of freedom is the way that the 24 is divided
between the two $E_8$ groups.

We thus expect that the $\SO(32)$ heterotic string determines a unique
value for $n$ whereas the $E_8\times E_8$ string requires further
specification of how the 24 is divided between the two factors. In
\cite{MV:F,MV:F2} it was argued that this split is $12-n$ and $12+n$
for the $E_8\times E_8$ string and it was conjectured that $n=4$ for
the $\SO(32)$ string.


\section{Explicit Models}  \label{s:exp}

$X$ is an elliptic fibration over $\HF n$ with $z$ the affine
coordinate of the generic $\P^1$ fibre of $\HF n$ and $w$ the
affine coordinate on the exceptional curve $C_0$ (i.e., the base). We
may then write the elliptic fibration in Weierstrass form
\begin{equation}
  y^2 = x^3 + a(w,z) x + b(w,z).	\label{eq:W}
\end{equation}
The fact that $X$ is a \CY\ space constrains $a$ and $b$ as
follows. The canonical class of $\HF n$ is
\begin{equation}
  K_{\HF n} = -2[C_0] - (2+n)[f],
\end{equation}
where $[C_0]$ is the class of $C_0$ and $[f]$ is the class of the
generic $\P^1$ fibre. Let $L$ be a line bundle such that $c_1(L)=-K_{\HF
n}$. Then $a$ must be a section of $L^{\otimes 4}$ and $b$ must be a
section of $L^{\otimes 6}$ \cite{Nak:Wei}.

To make direct contact with the heterotic string we look for ways in
which enhanced gauge groups can appear. It is generally believed that
this occurs in the type IIA string when ADE-like singularities are
acquired along a curve in $X$ 
\cite{me:flower,BSV:D-man,me:en3g,KMP:enhg}. In the first instance we
will be interested in the case 
that these groups are understood perturbatively from the heterotic
string. It was shown in \cite{me:en3g} that this corresponds to the
generic fibre of $X$ as a K3-fibration containing an ADE-like
singularity. 

The appearance of ADE-like singularities is well-understood in the
case of elliptic surfaces and the possibilities are listed in
\cite{MV:F2}. The $E_8\times E_8$ case is relatively straight-forward
as explained in \cite{MV:F2} and we quickly repeat it here as a guide
to the more subtle $\SO(32)$ case. Let us force $X$ to have an
$E_8\times E_8$ singularity in every generic K3 fibre. This requires
two ${\rm II}^*$ fibres in Kodaira's notation. This may be done by
specializing (\ref{eq:W}) to the form
\begin{equation}
  y^2 = x^3 + f(w)z^4 x + g_1(w)z^5 + g_2(w)z^6 + g_3(w)z^7.
		\label{eq:E8xE8}
\end{equation}
It then follows immediately from the \CY\ condition on $X$ above that
$f(w)$ is of degree 8 in $w$, $g_1(w)$ is of degree $12-n$, $g_2(w)$ is of
degree $12$ and $g_3(w)$ is of degree $12+n$.

At the $12-n$ zeros of $g_1(w)$ and the $12+n$ zeros of $g_3$ the
singularities of the K3 fibre get worse. This is interpreted in the
heterotic string language as follows \cite{MV:F2}. To get the full
$E_8\times E_8$ gauge group we go to an extreme case of the bundle $E$
which corresponds to 24 point-like instantons. Any smooth finite-sized
instanton will break part of the gauge group. Thus, these worse
singularities are simply the location of these 24 point-like
instantons. The fact that $12-n$ points worsen one of the ${\rm II}^*$
fibres and $12+n$ worsen the other corresponds to the identification
of $n$ as the way the 24 is split between the two $E_8$ factors as
stated above.

It is perhaps worth noting that some duality magic is already at work
here. The total instanton number of 24 was predicted by the \CY\
condition on $X$. One may also show, using the methods of \cite{MV:F2},
that this condition also imposes the \CY\ condition on the space the
heterotic string lives on, namely that it must be K3.

In order that $X$ be a smooth manifold, the fibration (\ref{eq:E8xE8})
must be blown-up (in addition to the blow-ups of the ${\rm II}^*$
fibres). In effect, the base $\HF n$ must be blown-up at 24
points. Given the results of the 
last section we may distinguish physically between blow-ups of the
elliptic fibres and blow-ups of the base. In the four-dimensional
picture of a type IIA string compactified on $X$, both blow-ups are
deformations of the K\"ahler form on $X$ and are thus moduli living in
four-dimensional {\em vector\/} multiplets. When we go to the
six-dimensional picture, the fibre must be shrunk down to zero
size. Thus the blow-ups of the fibre are not really moduli at all. The
blow-ups of the base are still moduli however. Since a tensor
multiplet in six dimensions contains a scalar and reduces to a vector
multiplet in four dimensions; and a vector multiplet in six dimensions
contains no scalars but also reduces to a vector multiplet in four
dimensions the interpretation is clear when viewing $X$ as an elliptic
fibration:
\begin{itemize}
  \item Requiring a blow-up in the fibre indicates the appearance of a
nonabelian gauge group.
  \item Requiring a blow-up of the base indicates a tensor multiplet.
\end{itemize}

Thus the compactification of the $E_8\times E_8$ string with zero
sized instantons contains generically no further enhancement of the gauge group
beyond what is know perturbatively, 
but will contain 24 massless scalars coming from tensor multiplets which
have no perturbative interpretation for the heterotic string
\cite{SW:6d}.
These extra moduli can be used to perform extremal transitions between
heterotic strings of different values of $n$ \cite{SW:6d,MV:F,CaFo:web}.

Now we will try to do the same analysis for the $\SO(32)$ string. 
Firstly we ask that the theory (with the instantons shrunk down to
points) has an unbroken perturbative gauge group of $\SO(32)$. This
implies that the generic K3 fibre has an ${\rm I\vphantom{II}}^*_{12}$
fibre itself.

Let us assume that this happens when $z=0$. This means that (see, for
example, \cite{Mir:fibr}) 
\begin{equation}
  \eqalign{a(w,z) &= z^2a_0(w,z),\cr
   b(w,z) &= z^3b_0(w,z),\cr
   \Delta = 4a(w,z)^3+27b(w,z)^2 &= z^{18}\Delta_0,\cr} \label{eq:I12}
\end{equation}
where $\Delta$ is the discriminant of the elliptic fibre and
$a_0(w,z)$, $b_0(w,z)$ and $\Delta_0$ are nonzero for $z=0$. We also
subject $a(w,z)$ and $b(w,z)$ to the condition above that $X$ be a
\CY.

Fixing $w$, we expect to obtain a two parameter family of K3 surfaces. This is
because our K3 surface has Picard number $20-2=18$. The problem is
that, unlike the $E_8\times E_8$ case, the space of solutions is not
connected. To fix which component of the moduli space is suitable for
the $\SO(32)$ heterotic string, we require a further constraint.

The constraint we have not yet used is the fact that the $\SO(32)$
heterotic string is built using the even self-dual Barnes--Wall
lattice $\Gamma_{16}$, as opposed to the root lattice of $\SO(32)$ which
is {\em not\/} self-dual. Given the duality between the Picard lattice
of the K3 surface and the set of winding/momenta modes of the
heterotic string, this translates into the statement that the Picard
lattice of the K3 surface, $S_w$, for a generic $w$, is self-dual. 

The Picard lattice of a surface as an elliptic fibration has been
analyzed in \cite{MirPer:ell}. The result we require is as
follows. Let the surface $S$ be an elliptic fibration and let $\Phi$
be the group of global sections, which we assume to be finite. Fix a
section of $S$ and let $R$ be the sublattice of $\Pic(S)$ generated by
components of fibres not meeting this section. We then have
\begin{equation}
  \disc(R) = |\Phi|^2\disc(\Pic(S)).
\end{equation}
In the case of an ${\rm I\vphantom{II}}^*_{12}$ fibre we know that $R$
is the root 
lattice of $\SO(32)$ and thus has discriminant 4. It follows that we
require $|\Phi|=2$, i.e., {\em $S_w$ must have two global sections as an
elliptic fibration}.

The Weierstrass form (\ref{eq:W}) has generically only one global
section --- at $x=y=\infty$. To force another global section we may
write it in the more restricted form
\begin{equation}
  y^2 = \left(x\vphantom{x^2}-\alpha\left(w,z\right)\right)
	\left(x^2+\alpha(w,z)x+\beta(w,z)\right),  \label{eq:W2}
\end{equation}
which gives a section along $x=\alpha(w,z)$, $y=0$.
From (\ref{eq:I12}) we see that $\alpha$ must have a zero of order 1 at
$z=0$ and $\beta$ must vanish to order 2.
Now when we demand that $\Delta$ has a zero of order 18 we obtain a
connected set of solutions as expected. The following represents a solution
sufficiently generic for our purposes:
\begin{equation}
  \eqalign{
    \alpha(w,z) &= Bz^4+Cz\cr
    \beta(w,z) &= Az^8-4BCz^5-2C^2z^2.\cr}
\end{equation}
The \CY\ condition then imposes that $A$ is a polynomial of degree $8+4n$ in
$w$, $B$ is of degree $4+2n$, and $C$ is of degree $4-n$. The
discriminant on $\HF n$ takes the form
\begin{equation}
  \Delta = z^{18}\left(A+2B^2\right)^2\left((4A-B^2)z^6
	-18BCz^3-9C^2\right).	\label{eq:Discr}
\end{equation}

Clearly from (\ref{eq:Discr}) we see the ${\rm I\vphantom{II}}^*_{12}$
fibres along $z=0$ generating the $\SO(32)$ gauge group. We also see
that interesting 
things are going to happen at the $8+4n$ zeros of
$A+2B^2$. Identifying these as the small instantons of the $\SO(32)$
heterotic string, we see independent evidence that we require $n=4$. In
this case there will be 24 small instantons --- precisely what is
required for the $\SO(32)$ gauge symmetry to appear. From now on we
will assume that $n=4$.

A remarkable agreement with the results of \cite{W:small-i} occurs
when we ask exactly what happens at the zeros of $A+2B^2$. Consider a
zero of $A+2B^2$ of order $k$ in $w$. This will correspond to $k$
point-like instantons coalescing. The generic case will be $k=1$. The
zero of the discriminant (\ref{eq:Discr}) will be of order $2k$ in $w$
for any generic value of $z$. The terms $a(w,z)$ and $b(w,z)$ are
generically nonzero. Thus according to \cite{Mir:fibr}, we have a
curve of ${\rm I\vphantom{II}}_{2k}$ fibres. As mentioned earlier,
blow-ups in such a fibre correspond to extra gauge groups
appearing. That is, each small instanton is generating a new
nonperturbative contribution to the gauge group. At first sight,
according to \cite{MV:F2}, the gauge group associated to ${\rm
I}_{2k}$ is $\SU(2k)$. There is a subtlety which changes this however.

In the analysis of \cite{me:en3g}, where an enhanced gauge group
appeared because of a degeneration of every generic K3 fibre, care was
taken to restrict to the case where the Picard group of the K3 fibre
was {\em monodromy invariant\/} as one moved around the base
$\P^1$. In the case at hand, it turns out that the rational curves
within the blow-up of the ${\rm I}_{2k}$ fibre are not monodromy
invariant as we move around the curve within $\HF 4$ of fixed $w$.
To obtain the enhanced gauge group that really appears, we require the
monodromy invariant part of the apparent local gauge group over each
point. The action of the monodromy on the blown-up fibre can be
translated into an action on the Dynkin diagram of the simply-laced
local gauge group. The required group is then the subgroup invariant
under this outer automorphism. We can thus obtain non-simply-laced
gauge groups in the type II setting. This is fairly analogous to the
way such groups appeared in the heterotic string \cite{CP:ao}.

\iffigs
\begin{figure}
  \centerline{\epsfxsize=13cm\epsfbox{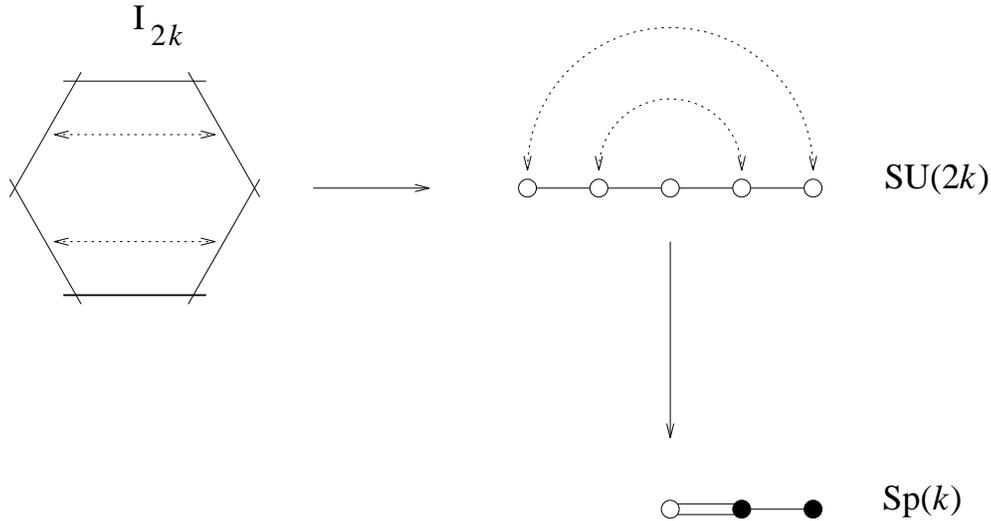}}
  \caption{The generation of the gauge group $\Sp(k)$.}
  \label{fig:Sp}
\end{figure}
\fi

In the case at hand, there is a $\Z_2$ monodromy group which acts on
the ${\rm I}_{2k}$ fibre as shown in 
figure \ref{fig:Sp} (for the case $k=3$). This was demonstrated in
\cite{Mir:fibr}. 
Results for finding the invariant part of Lie algebras under outer
automorphisms may be found in \cite{FulHar:rep} for example.
We see that the gauge
group which appears when $k$ instantons coalesce is $\Sp(k)$ in
complete agreement with \cite{W:small-i}.


\section{Hypermultiplets}   \label{s:hyper}

In \cite{W:small-i} it was predicted that one should also obtain
hypermultiplets in a $(\mbox{\bf 32},\mbox{\boldmath {\bf 2}$k$})$
representation of $\SO(32)\times\Sp(k)$ when small instantons
appear. It was suggested in \cite{MV:F2} that the appearance of
hypermultiplets was connected somehow to the collision of loci of bad
elliptic fibres within the base $\HF n$. We will attempt to go some
way to justifying these statements here.

First recall the ``wrapping $p$-brane'' of \cite{Str:con} and the way
in which it applied to enhanced gauge symmetries of a type IIA string
on a K3 surface. When rational curves in the K3 surface shrink down to
zero size, extra massless particles appear whose charges fill out the
root lattice of the resulting nonabelian group. The charges are
dictated by their interaction with the $U(1)$ vector fields which in
turn can be measured by how their mass increases as moduli within
these vector supermultiplets are switched on to Higgs away the
enhanced gauge symmetry. The mass is given by the area which in turn
is given by the K\"ahler form's value on the homology class of the rational
curve. That is, the set of elements in $H_2$ of the rational curve
classes around 
which the 2-branes wrap form the root lattice of the enhanced gauge
group.

For the type IIA string compactified on $X$, the above picture happens
along a whole curve of ADE-like singularities to produce an enhanced
gauge group. At points along this curve, the singularities may be worse. At
these points there can be extra classes of rational curves. These need
not correspond to extra vector moduli. That is, blowing-up the curve
of ADE-like singularities may well resolve the ``bad point'' on the
curve too. However, even in this case the extra classes of rational
curves can produce more massless states when the gauge group is
enhanced.

From what we have said, the homology classes of these extra curves with
respect to the homology classes of the rational curves which appear
when blowing up the curve of singularities should give the charges of
these new states with respect to the Cartan subgroup of the gauge
group. That is, we expect these new homology classes to form the {\em
weight lattice\/} of the new representation. In this way, we expect
the massless hypermultiplets to appear.

Let us apply this approach to our $\SO(32)\times\Sp(k)$ problem. In
this case we have a curve of ${\rm I\vphantom{II}}^*_{12}$ fibres
intersecting a curve of ${\rm I}_{2k}$
fibres transversely at one point. It is at this point of intersection
that the new classes can appear. This problem was analyzed in 
\cite{Mir:fibr,Gross:E3:2}. The result is as follows. Label the homology
classes of the exceptional curves in these fibres in terms of the Lie
algebras they represent as
\begin{equation}
\setlength{\unitlength}{0.008in}%
\begin{picture}(635,105)(70,680)
\thicklines
\put( 80,780){\circle{10}}
\put( 80,700){\circle{10}}
\put(120,740){\circle{10}}
\put(170,740){\circle{10}}
\put(220,740){\circle{10}}
\put(300,740){\circle{10}}
\put(350,740){\circle{10}}
\put(480,740){\circle{10}}
\put(530,740){\circle*{10}}
\put(580,740){\circle*{10}}
\put(630,740){\circle*{10}}
\put(700,740){\circle*{10}}
\put(125,740){\line( 1, 0){ 40}}
\put(175,740){\line( 1, 0){ 40}}
\put(305,740){\line( 1, 0){ 40}}
\multiput(225,740)(8.23529,0.00000){9}{\line( 1, 0){  4.118}}
\put( 84,776){\line( 1,-1){ 31.5}}
\put( 84,704){\line( 1, 1){ 31.5}}
\put(535,740){\line( 1, 0){ 40}}
\put(585,740){\line( 1, 0){ 40}}
\multiput(635,740)(8.00000,0.00000){8}{\line( 1, 0){  4.000}}
\put(485,743){\line( 1, 0){ 42}}
\put(485,738){\line( 1, 0){ 42}}
\put( 70,760){\makebox(0,0)[lb]{$\alpha_1$}}
\put( 75,680){\makebox(0,0)[lb]{$\alpha_2$}}
\put(120,720){\makebox(0,0)[lb]{$\alpha_3$}}
\put(165,720){\makebox(0,0)[lb]{$\alpha_4$}}
\put(215,720){\makebox(0,0)[lb]{$\alpha_5$}}
\put(295,720){\makebox(0,0)[lb]{$\alpha_{15}$}}
\put(345,720){\makebox(0,0)[lb]{$\alpha_{16}$}}
\put(475,715){\makebox(0,0)[lb]{$\beta_1$}}
\put(525,715){\makebox(0,0)[lb]{$\beta_2$}}
\put(575,715){\makebox(0,0)[lb]{$\beta_3$}}
\put(625,715){\makebox(0,0)[lb]{$\beta_4$}}
\put(695,715){\makebox(0,0)[lb]{$\beta_k$}}
\end{picture}
\end{equation}

When the curves of ${\rm I\vphantom{II}}^*_{12}$ and ${\rm I}_{2k}$ fibres 
collide transversely, the result at the point of collision according to
\cite{Mir:fibr} is an ${\rm I\vphantom{II}}^*_{12+k}$ fibre. We label
these classes as
\begin{equation}
\setlength{\unitlength}{0.008in}%
\begin{picture}(635,105)(-50,680)
\thicklines
\put( 80,780){\circle{10}}
\put( 80,700){\circle{10}}
\put(120,740){\circle{10}}
\put(170,740){\circle{10}}
\put(220,740){\circle{10}}
\put(300,740){\circle{10}}
\put(350,740){\circle{10}}
\put(125,740){\line( 1, 0){ 40}}
\put(175,740){\line( 1, 0){ 40}}
\put(305,740){\line( 1, 0){ 40}}
\multiput(225,740)(8.23529,0.00000){9}{\line( 1, 0){  4.118}}
\put( 84,776){\line( 1,-1){ 31.5}}
\put( 84,704){\line( 1, 1){ 31.5}}
\put( 70,760){\makebox(0,0)[lb]{$\gamma_1$}}
\put( 75,680){\makebox(0,0)[lb]{$\gamma_2$}}
\put(120,720){\makebox(0,0)[lb]{$\gamma_3$}}
\put(165,720){\makebox(0,0)[lb]{$\gamma_4$}}
\put(215,720){\makebox(0,0)[lb]{$\gamma_5$}}
\put(290,720){\makebox(0,0)[lb]{$\gamma_{15+k}$}}
\put(345,720){\makebox(0,0)[lb]{$\gamma_{16+k}$}}
\end{picture}
\end{equation}
Now we see from \cite{Gross:E3:2} that, among the relations, we have
\begin{equation}
  \eqalign{\alpha_1&=\gamma_1\cr
  \alpha_2&=\gamma_2\cr
  \beta_1&=\gamma_1+\gamma_2+2\gamma_3.\cr}
\end{equation}
That is,
\begin{equation}
  \gamma_3 = -\ff12\alpha_1-\ff12\alpha_2+\ff12\beta_1.  \label{eq:w1}
\end{equation}
Now $-\ff12\alpha_1-\ff12\alpha_2+\ff12\beta_1$ is a weight of
the $(\mbox{\bf 32},\mbox{\boldmath {\bf 2}$k$})$
representation. Furthermore we may obtain the rest of the weights of
this representation by adding and subtracting
integer multiples of roots. Thus we see that the extra classes
introduced by the collision can produce the required extra
hypermultiplets. The other $\gamma_i$'s do not introduce any further
new weights.

We cannot claim to have shown precisely the existence of every
element of the weight lattice here since we need to show that the
2-brane solitons wrap precisely once around each required class on
$H_2$. One might try to use some kind of D-brane analysis to prove
this. What we can claim however is that if the gauge symmetry exists
then we must get a valid weight system for the Lie algebra, and then 
the weight system of $(\mbox{\bf 32},\mbox{\boldmath {\bf 2}$k$})$ is
the minimal such system that contains the rational curve classes we
find.

Note we may apply the same techniques to the $E_8\times E_8$ case as
suggested in \cite{MV:F2}. Suppose we consider the generic case of an
$E_7$ gauge group. This involves a curve of ${\rm III}^*$ fibres. One
can also see that ${\rm I}_1$ fibres will cross (non-transversely) this
curve at $8-n$ points \cite{MV:F2}. Analyzing one of the points of
intersection, one may show that a rational curve lives there in one of
the weights of the {\bf 56} of $E_7$. (We suppress a lengthy algebraic
calculation). Note that a hypermultiplet 
contains both the representation and its complex conjugate of
fields. Since the {\bf 56} is a quaternionic (or pseudo-real)
representation we should really require two to form a single
hypermultiplet. We thus expect $\ff12(8-n)$ {\bf 56} hypermultiplets
in agreement with the index formula \cite{KV:N=2}.

What about the generic case of a curve of ${\rm IV}^*$ fibres? At first
sight, this case would appear to give an $E_6$ gauge group. There are 
$12-2n$ points along this curve where ${\rm I}_1$ fibres cross. It
turns out that around these collisions there is a $\Z_2$ monodromy
acting on the ${\rm IV}^*$ fibre. If $n=6$, or the $12-2n$ roots pair
up in the right way to kill the monodromy, we will have an $E_6$ gauge group.
Otherwise we claim that the gauge
group is actually $F_4$ as this is the invariant part of $E_6$ divided
by a $\Z_2$ outer automorphism.\footnote{This
monodromy was overlooked in 
\cite{MV:F2}. A gauge group of $F_4$ for the generic $n=5$ case is
also in agreement with the heterotic string picture. See, for example,
\cite{CaFo:web}.} One might expect
to see {\bf 26}'s of $F_4$ appearing 
at these crossings but since the weights of the {\bf 26} are actually
a subset of the roots, we have no way of deducing this with the methods
introduced in this section.


\section{Discussion}	\label{s:disc}

The excellent agreement for the analysis of the $\SO(32)$ heterotic
string between the methods shown above and those of \cite{W:small-i}
appears to be a very strong test of string duality. 
We also seem to be
seeing peculiar relationships between the algebraic geometry of a
\CY\ and the properties of vector bundles on a K3 surface. 
Firstly there are the properties of the enhanced gauge groups
themselves, as seen above and in \cite{me:en3g,MV:F,CaFo:web,MV:F2}
for example. Secondly we have seen how the appearance of rational
curves may be linked to massless hypermultiplets.

It is worth emphasizing a difference between the behaviour of generic
point-like instantons on the context of the $E_8\times E_8$ string and
the $\SO(32)$ string. For $E_8\times E_8$, the small instantons
have the following properties \cite{SW:6d}
\begin{enumerate}
  \item There is no nonperturbative enhancement of the gauge group.
  \item There is a massless tensor multiplet for each isolated
point-like instanton.
\end{enumerate}
For the $\SO(32)$ case we have \cite{W:small-i}
\begin{enumerate}
  \item There is an $\Sp(1)$ contribution to the nonperturbative part
of the gauge group for each isolated point-like instanton.
  \item There are no massless tensor multiplets.
  \item When point-like instantons coalesce, the gauge group enhances
further to $\Sp(k)$.
\end{enumerate}
What happens when point-like instantons of the $E_8\times E_8$ string
coalesce? The elliptic fibration picture shows that blow-ups in the
base are required just like the isolated case, and the total number of
blow-ups required is the same as the isolated case. The difference is
that in this case the 
exceptional divisors intersect. This makes it look very analogous to
the coalesced instantons in the $\SO(32)$ string, except that blow-ups
are in the base rather than the fibre. We therefore suggest the
following interpretation. When $k$ point-like instantons of the $E_8\times
E_8$ string coalesce, extra massless tensor multiplets appear. The
tensor multiplets associated to one point have mutual interactions
however and expectation values can only be given to $k$ of the them
simultaneously.  


\section*{Acknowledgements}

We would like to thank V.~Kaplunovsky,
D.~Morrison, R.~Plesser and H.~Tye for useful conversations.
The work of the authors is supported by grants from
the National Science Foundation.


\end{document}